\documentclass[journal=jpccck,manuscript=article]{achemso}

\usepackage[version=3]{mhchem} 

\author{Raphael M. Tromer}
\affiliation[State University of Campinas]
{Applied Physics Department, State University of Campinas, Campinas, SP, 13083-970, Brazil}
\alsoaffiliation[State University of Campinas]
{Center for Computational Engineering and Sciences, State University of Campinas, Campinas, SP, 13083-970, Brazil}
\author{Leonardo D. Machado}
\affiliation[Federal University of Natal]
{Departamento de F\'isica Te\'orica e Experimental, Universidade Federal do Rio Grande do Norte, Natal, RN, 59072-970, Brazil.}
\author{Cristiano F. Woellner}
\affiliation[Federal University of Parana]
{Physics Department, Federal University of Parana, UFPR, Curitiba, PR, 81531-980, Brazil}
\author{Douglas S. Galvao}
\email{galvao@ifi.unicamp.br}
\affiliation[State University of Campinas]
{Applied Physics Department, State University of Campinas, Campinas, SP, 13083-970, Brazil}

\title{Thiophene-Tetrathia-Annulene monolayer (TTA-2D): A new 2D semiconductor material with indirect bandgap}

\usepackage{xcolor}
\begin{document}


\begin{abstract}
We propose a new 2D semiconductor material (TTA-2D) based on the molecular structure of Thiophene-Tetrathia-Annulene (TTA). The TTA-2D structural, electronic, and optical properties were investigated using \textit{ab initio} methods. Our results show that TTA-2D is a small indirect bandgap semiconductor ($0.6$ eV). A semiconductor-metal transition can be induced by applying a uniaxial strain. Our results also show that TTA-2D is thermally stable up to $T=1000$~K. TTA-2D absorbs in a large spectral range, from infrared to ultraviolet regions. Values of refractive index and reflectivity show that TTA-2D reflects only $10\%$ of the incident light in the visible region. These results suggest that TTA-2D is a promising material for solar cell applications.
\end{abstract}

\section{Introduction}

Semiconductor materials have been largely used in technological applications for decades, and most common semiconductor devices are based on silicon and germanium \cite{BARTH_2010,BENT_2002,MD_2003,Pietro_2009,zheng_26mar2020}. In the future, the tendency is that devices will be comprised of new materials, such as graphene. Graphene was separated from bulk graphite through a  mechanical exfoliation method in 2004 \cite{Novoselov2004,Novoselov2005} and, because of its unique and excellent electronic properties, it has received special attention as a 
potential material for flexible electronic devices \cite{Gwon2011}. But graphene is a zero bandgap semiconductor, which prevents its use in some digital electronics applications. In order to address this limitation, many approaches have been tried to open the bandgap of graphene, including impurity adsorption, chemical doping, strain application \cite{Choi_2010}, among others. 

In the last years, many non-zero bandgap 2D materials similar to graphene have been investigated and/or synthesized, such as hexagonal boron nitride \cite{BN_synt,BN_synt2}, nitrogenated holey graphene \cite{Mahmood2015,tromer_2017}, and hematene \cite{hematene2018}. Other possible systems include structures composed of porphyrin molecules, which have been investigated recently. Porphyrin molecules serve as structural blocks that can be bonded together to build much larger systems. There are many polymerization methods used for building large blocks of 2D polymers from porphyrin organic molecules \cite{Appli_1,Appli_2,Appli_4,Zhong1379}. The first 2D system based on porphyrin phthalocyanine-iron molecules was synthesized in 2010 by Abel and co-authors \cite{Abel2010}. An experimental method to synthesize a 2D porphyrin covalent organic framework with control of the pore sizes was proposed by Jiang et al. \cite{Jiang_2011}. Recently, the electronic and magnetic properties of these systems were investigated using \textit{ab initio} methods \cite{por_ferromagnetism,por_electronic}. More recently \cite{Tromer2020}, a new 2D semiconductor based on diboron-porphyrin (a molecule already synthesized \cite{Brothers2011}) was proposed. 

Besides porphyrin, different types of organic molecules have been investigated to serve as structural blocks for building large 2D systems \cite{2D_pi}. Some of these structures used thiophene (sulfur) \cite{tio_mol} instead of pyrrol (nitrogen) units. They can be very similar to porphyrin, because both exhibit well delocalized $\pi$ molecular orbitals \cite{Kano_1997,posli_2018}. Large thiophene-based systems have been used in electronic, optical, and magnetic applications such as field-effect transistors (FETs), light-emitting diodes (LEDs) and solar cells \cite{NIELSEN_2013,MEHMOOD_2016}. Some hybrid systems combining pyrrol and thiophene have also been proposed \cite{tiopor}. In 2013, the  world's first supramolecular thiophene nanosheet \cite{zheng_26mar2020} was reported. This 2D single-layered structure is composed of linked thiophene molecules and exhibits very promising properties for applications in electronic devices. 
 
 In this work, we propose a new thiophene based 2D system, based on the Thiophene-Tetrathia-Annulene (TTA) molecule (Fig. \ref{fig:structure}), which was synthesized by Cava \textit{et al.} \cite{TTA}. The atomic arrangement in this structure is similar to that of porphyrin molecules. TTA exhibits electronic and optical properties suitable for applications in optoelectronics \cite{TTA2,TTA3}. We have investigated the structural stability, electronic, optical, and mechanical properties of single-layer TTA-based crystals (TTA-2D) using density functional theory (DFT) methods.
 The unit cell we considered was generated by replacing TTA carbon-hydrogen bonds by carbon-carbon bonds (Fig. \ref{fig:structure}). Our results show that TTA-2D is a small indirect bandgap semiconductor, with a formation energy similar to that of graphene. It exhibits high thermal stability for temperatures up to $1000$ K. We also determined the effect induced on the electronic and optical properties by an applied uniaxial strain.

\section{Methodology}
In this work, we carried out density functional theory (DFT) calculations using the SIESTA software \cite{Soler_2002}, which uses a basis set of localized atomic orbitals. The calculations were carried within the generalized gradient approximation (GGA) using the exchange-correlation functional by Perdew-Burke-Ernzenhof (PBE) \cite{Perdew1996}. The core electrons are described by norm-conserving Troullier-Martins pseudopotentials, and the basis set by double-zeta plus polarization functions (DZP).
The kinetic energy cut-off was $250$ Ry, and the reciprocal space was sampled using a $4\times 4\times 1$ k-point grid. In order to prevent spurious interactions between images located at the perpendicular direction to the plane, we added a $20$ \AA~ vacuum along the non-planar direction where the lattice cell is kept fixed during the simulation. The lattice vectors in the plane direction and ionic positions were fully relaxed until the maximum force on each atom was less than $0.005$ eV/\AA.
The structural stability was tested considering the phonon dispersion analysis (no negative frequencies). The TTA-2D unit cell contains 26 atoms (22 carbon and 4 sulfur atoms).

In order to obtain stress-strain curves, we applied a uniaxial strain along the TTA-2D $X$ and $Y$ directions and then determined the stress response. However, prior to obtaining Young's modulus values, it is important to realize that SIESTA uses the volume of the unit cell to calculate the stress. Hence, we multiply the stress values provided by SIESTA (in units of eV/A$^3$), by a factor (L$_z$/L$_C$), where $L_z$ is the length of the vacuum region and $L_C$ is the thickness of TTA-2D, in \AA~units. We assumed a thickness of $3.33$ \AA \ for TTA-2D, equal to that of graphene \cite{Felix2018,felix_2019,Felipe}. Next, a linear fit was performed in the region with a strain between $0$ and $1\%$, and the angular coefficient of this fit allow us to the Young's modulus values.

We also investigated the thermal structural stability of TTA-2D, from 0 K up to 1000K. We carried out \textit{ab initio} molecular dynamics (AIMD) simulations within the NVT ensemble and used a supercell with 104 atoms. The unit cell was replicated 2 times in the $X$, 2 times in the $Y$, and once in the $Z$ direction. 

In order to perform optical calculations, we applied an external electric field of magnitude $1.0$ V/\AA. This is a typical value for an organic system, allowing the optical stimulation to occur without significant distortion in the structure (linear regime) \cite{Fadaie2016}. We considered polarization along one specific direction ($X$ or $Y$). We do not discuss results along the $Z$ direction, because the absorption is much smaller in the $Z$ than in the $X$ or $Y$ directions.     
The complex dielectric function is defined by $\epsilon =\epsilon_1+i\epsilon_2$, where the imaginary part, $\epsilon_2$, can be extracted from direct interband transitions via Fermi's golden rule,
\begin{equation}
\epsilon_2(\omega)=\frac{4\pi^2}{\Omega\omega^2}\displaystyle\sum_{i\in \mathrm{VB},j\in \mathrm{CB}}\displaystyle\sum_{k}W_k|\rho_{ij}|^2\delta	(\epsilon_{kj}-\epsilon_{ki}-\omega).
\end{equation}
In the above expression, VB and CB are valence and conduction bands, $\omega$ is the photon frequency, $\rho_{ij}$ is the dipole transition matrix element, and $\Omega$ is the unit cell volume.

The real part $\epsilon_1$ is obtained through the Kramers-Kronig relation:
\begin{equation}
\epsilon_1(\omega)=1+\frac{1}{\pi}P\displaystyle\int_{0}^{\infty}d\omega'\frac{\omega'\epsilon_2(\omega')}{\omega'^2-\omega^2},
\end{equation}
where $P$ denotes the principal value.

Once the complex dielectric function is obtained, any optical properties of interest -- such as the absorption coefficient $\alpha$, the reflectivity $R$, and the refractive index $\eta$ -- can also be determined:

\begin{equation}
\alpha (\omega )=\sqrt{2}\omega\bigg[(\epsilon_1^2(\omega)+\epsilon_2^2(\omega))^{1/2}-\epsilon_1(\omega)\bigg ]^{1/2},
\end{equation}
\begin{equation}
R(\omega)=\bigg [\frac{(\epsilon_1(\omega)+i\epsilon_2(\omega))^{1/2}-1}{(\epsilon_1(\omega)+i\epsilon_2(\omega))^{1/2}+1}\bigg ]^2 ,
\end{equation}
\begin{equation}
\eta(\omega)= \frac{1}{\sqrt{2}} \bigg [(\epsilon_1^2(\omega)+\epsilon_2^2(\omega))^{1/2}+\epsilon_1(\omega)\bigg ]^{2}.
\end{equation}

Optical transitions are directly related to the band structure of materials. For instance, the first optical transition is also the most relevant, and it depends on the value of the bandgap. Therefore, in order to describe the optical properties of TTA-2D with some accuracy, it is very important to calculate the bandgap accurately. As mentioned early, we use the GGA-PBE approximation to represent the exchange and correlation functional. In the literature, it is established that this approximation does not reproduce correctly the gap energies of semiconductor materials  \cite{Johnson_1998}. Other approximations, such as the HSE06 hybrid functional, yield more accurate bandgap values \cite{Kishore2017}. Therefore, we used the HSE06 functional implemented in the gaussian16 software to obtain the bandgap energy of TTA-2D with more accuracy. 
Then, we used the HSE06 bandgap energy to correct our optical calculations, using the scissor operator available in the SIESTA code. In this approach, a shift in the unoccupied states is performed using the following expression,
\begin{equation}
\mathrm{scissor}=E_{gap}^{\mathrm{HSE06}}-E_{gap}^{\mathrm{PBE}}.
\label{scissor}
\end{equation}
Other articles that perform optical calculations with GGA-PBE have used this method to obtain corrected optical spectra. The results obtained agree with more accurate methods, such as the GW approximation, and with experimental results \cite{Nayebi2016,Ljungberg2017,Kolos2019}. 

We analyzed the influence of an externally  applied uniaxial strain (along the X and Y directions) on the electronic and optical properties. We considered strain values from 1 up to 10\%.

\section{Results}
\subsection{Structural Stability}
\begin{figure}[]
\begin{center}
\includegraphics[width=0.4\linewidth]{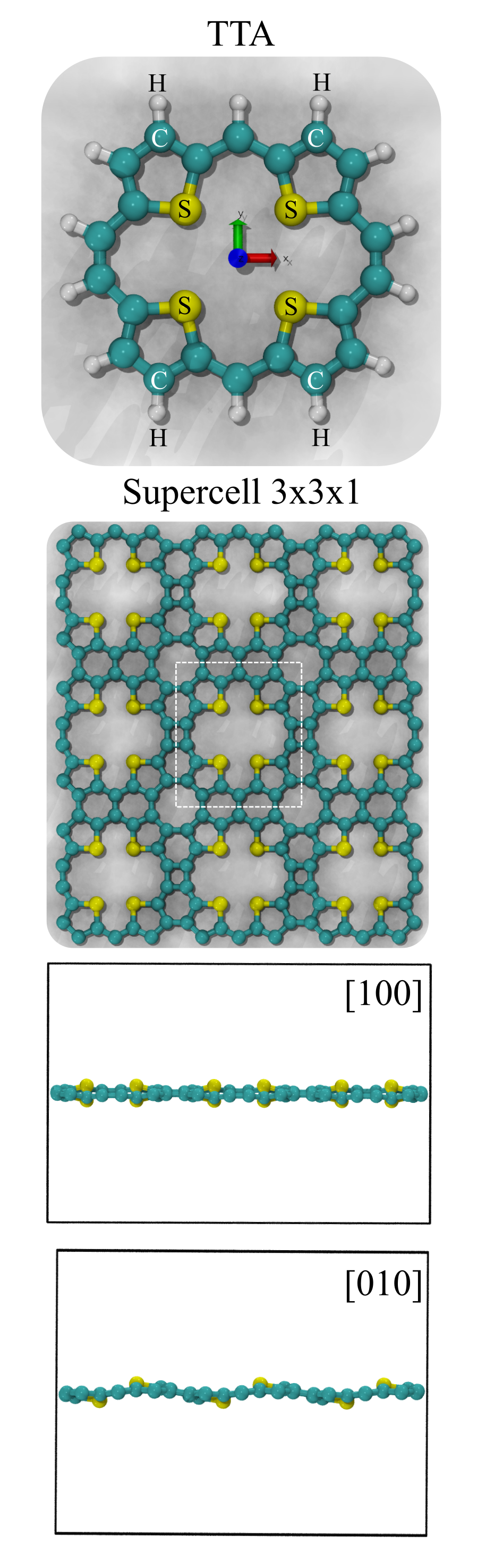}
\caption{Top: Atomic arrangement of the TTA molecule. The other images display the supercell of the proposed TTA-2D structure after optimization, viewed along different crystal axes. The dashed white line represents the orthorhombic unit cell with lattice vectors ${v}_x=8.74$~\AA~and ${v}_y=10.02$~\AA. In this figure, the unit cell was replicated $3 \times 3 \times 1$ times. Carbon (C) atoms are in grey, sulfur (S) atoms are in yellow and hydrogen (H) atoms are in white. }
\label{fig:structure}
\end{center}
\end{figure}

In Figure \ref{fig:structure} we present the TTA molecules and the TTA-2D optimized structure along different views. The minimum unit cell contains $26$ atoms and it is indicated by square lines and replicated along $X$ and $Y$ directions by $3\times 3 \times 1$. The orthorhombic crystal cell has vectors $v_x=8.74$~\AA~ and $v_y=10.02$~\AA~ that coincides with the plane of TTA-2D, while the other vector $v_z=20.0$~\AA~is perpendicular to the plane. We can see from Figure \ref{fig:structure} that the TTA-2D is not perfectly planar, and the out-of-plane distortions can be attributed to sulfur stereochemical effects. The main geometrical features of TTA-2D and the TTA molecule differ only by small distortions. The enthalpy of formation of TTA-2D is $-8.6$~eV/atom, which is comparable to other 2D systems in the literature such as graphene ($-8.8$~eV/atom) \cite{Wang2015}.    

\begin{figure}[]
\begin{center}
\includegraphics[width=1.0\linewidth]{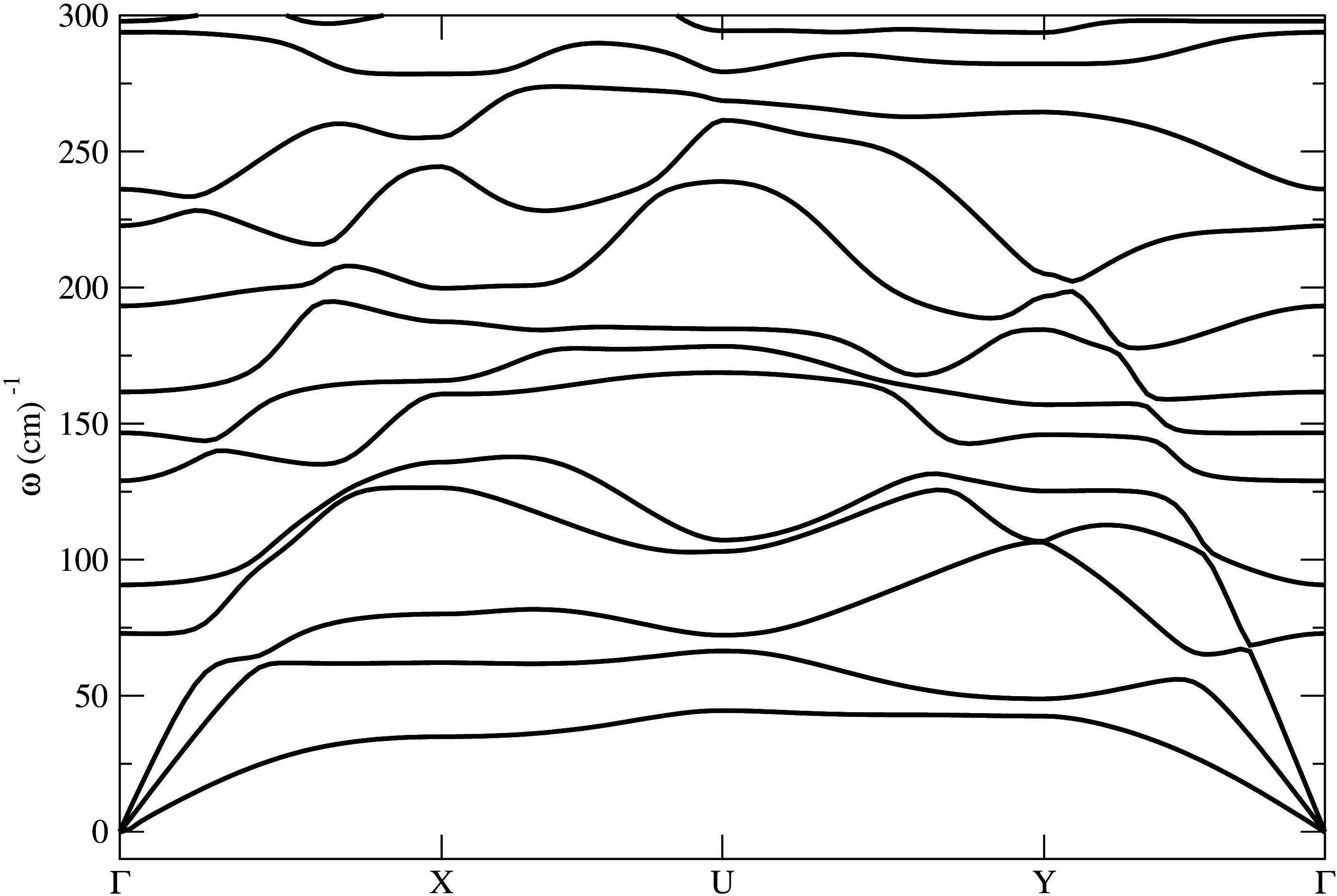}
\caption{Phonon dispersion calculation results for TTA-2D, considering a path corresponding to the symmetry points of an orthorhombic crystal cell.}
\label{fig:dispersao}
\end{center}
\end{figure}

To test the TTA-2D structural stability, we performed a phonon dispersion calculation for the optimized structure shown in Figure \ref{fig:structure}. In Figure \ref{fig:dispersao} we present the phonon dispersion results for the path corresponding to the orthorhombic unit cell. It is clear that no negative frequencies occur, thus corroborating the TTA-2D structural stability at $T=0$~K. Near the $\Gamma$ point there are three acoustic branches that correspond to two modes in-plane and one out-of-plane, as expected in 2D systems. The lowest optical branch occurs at $74$ cm$^{-1}$. The vibrations are not symmetric, the path $\Gamma$ to $U$ passing through $X$ are quite different for the same path but passing through $Y$. Therefore, the thermal transport in the TTA-2D must be anisotropic.

The TTA-2D Young's modulus values were obtained from the angular coefficient through a linear fit performed in the region with a strain between $0$ and $1\%$ (more details in the Supplementary Materials). The obtained Young's modulus values were $327.1$ and $476.1$ GPa for the $X$ and $Y$ directions, respectively. These values are larger than those obtained in other 2D systems with sulfur such as MoS2 ($270$~GPa) \cite{mos2}.

\begin{figure}[]
\begin{center}
\includegraphics[width=0.83\linewidth]{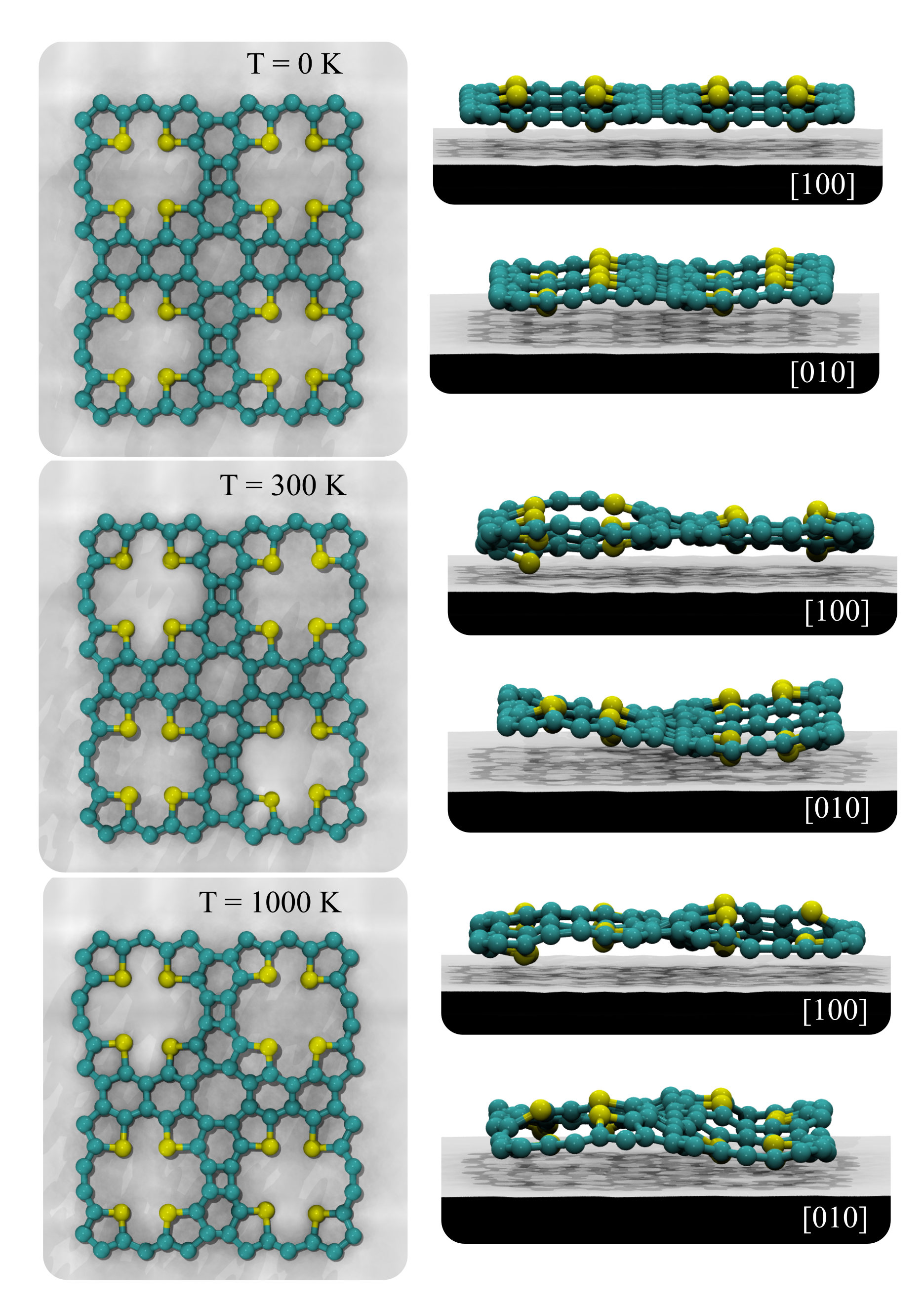}
\caption{Representative snapshots from AIMD simulations at different temperatures. The images show the TTA-2D structure after $2$~ps of the simulation time.}
\label{fig:md}
\end{center}
\end{figure}

We also investigated TTA-2D thermal stability. We carried out AIMD during $2$~ps with $dt=1$~fs within a NVE ensemble for the TTA-2D $2 \times 2\times 1$ supercells. In Figure \ref{fig:md} we present representative AIMD snapshots at different temperatures. We can see that even at $T=1000$~K the TTA-2D structural integrity remains, the major thermal effects are to increase the out-of-plane movements. These results indicate that TTA-2D can be structurally stable, even at high temperatures. These results are important because one of the synthetic approach to such structures \cite{Zhong1379} requires high temperatures. Thus, large TTA-2D fragments might be created using these techniques.

\section{Electronic Analysis}

\begin{figure}[]
\begin{center}
\includegraphics[width=0.83\linewidth]{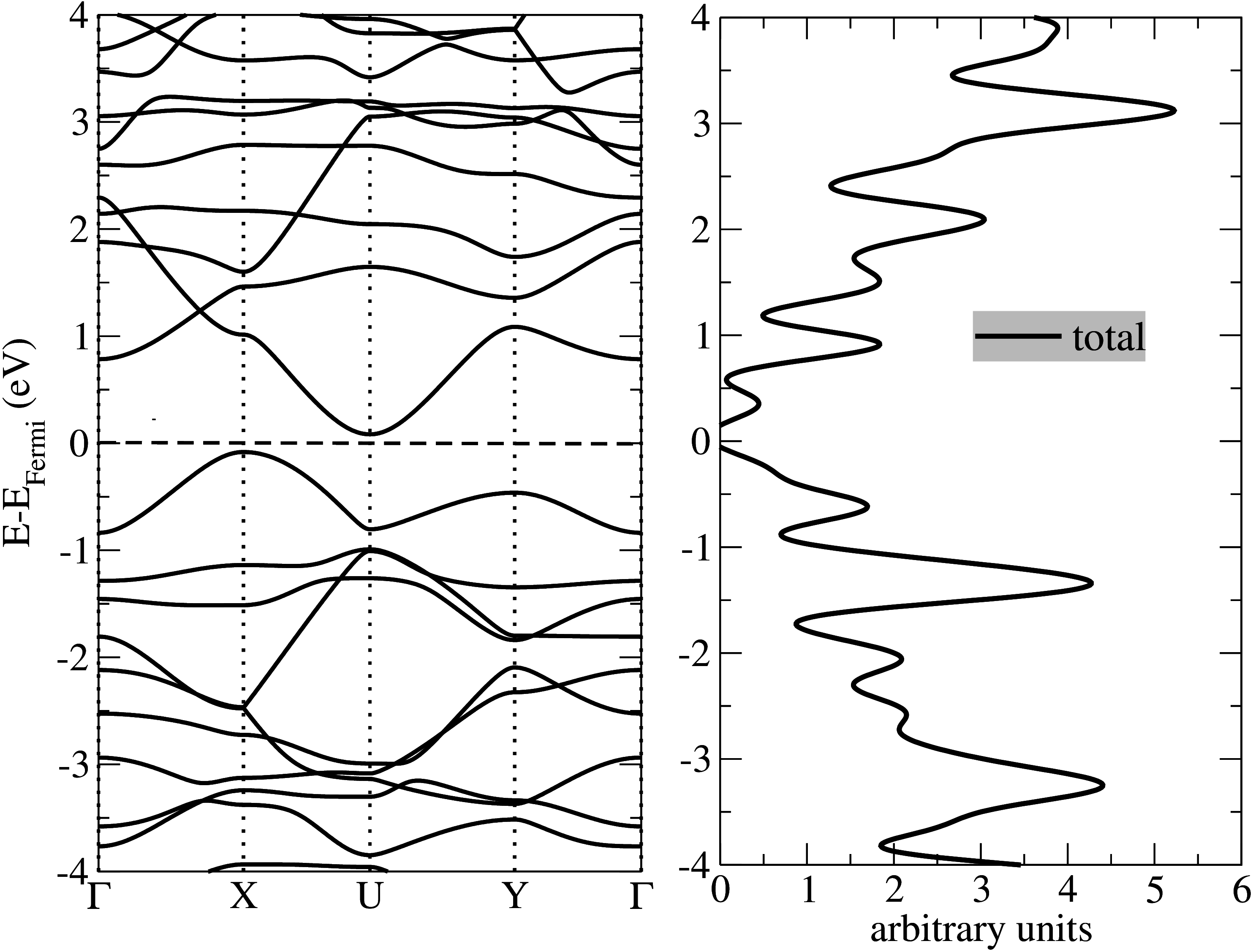}
\caption{TTA-2D electronic band structure and the corresponding total density of states (DOS).  \label{fig:banpdos}}
\end{center}
\end{figure}

In Figure \ref{fig:banpdos} we present the TTA-2D electronic band structure and the corresponding total density of states (DOS).
From the band structure, we observe that TTA-2D is a semiconductor material with a small indirect energy bandgap of $\approx 0.2$ eV, from $X$ to $U$ symmetry points. A more precise calculation using the functional HSE06 was also performed to obtain the bandgap value ($0.6$ eV). The TTA-2D valence bands are predominantly composed of atomic orbitals 2p$_z$ of carbon and 3p$_z$ of sulfur atoms. The other atomic orbitals slightly contribute to the valence bands. In contrast, the conduction bands are mainly composed of carbon 2p$_z$ orbitals. This characteristic where the atomic orbital p$_z$ has a major contribution to conduction bands are typical carbon-based 2D systems \cite{lilika}. We ran some calculation tests with spin polarization and we did not observe any magnetic moment.

\begin{figure}[]
\begin{center}
\includegraphics[width=0.5\linewidth]{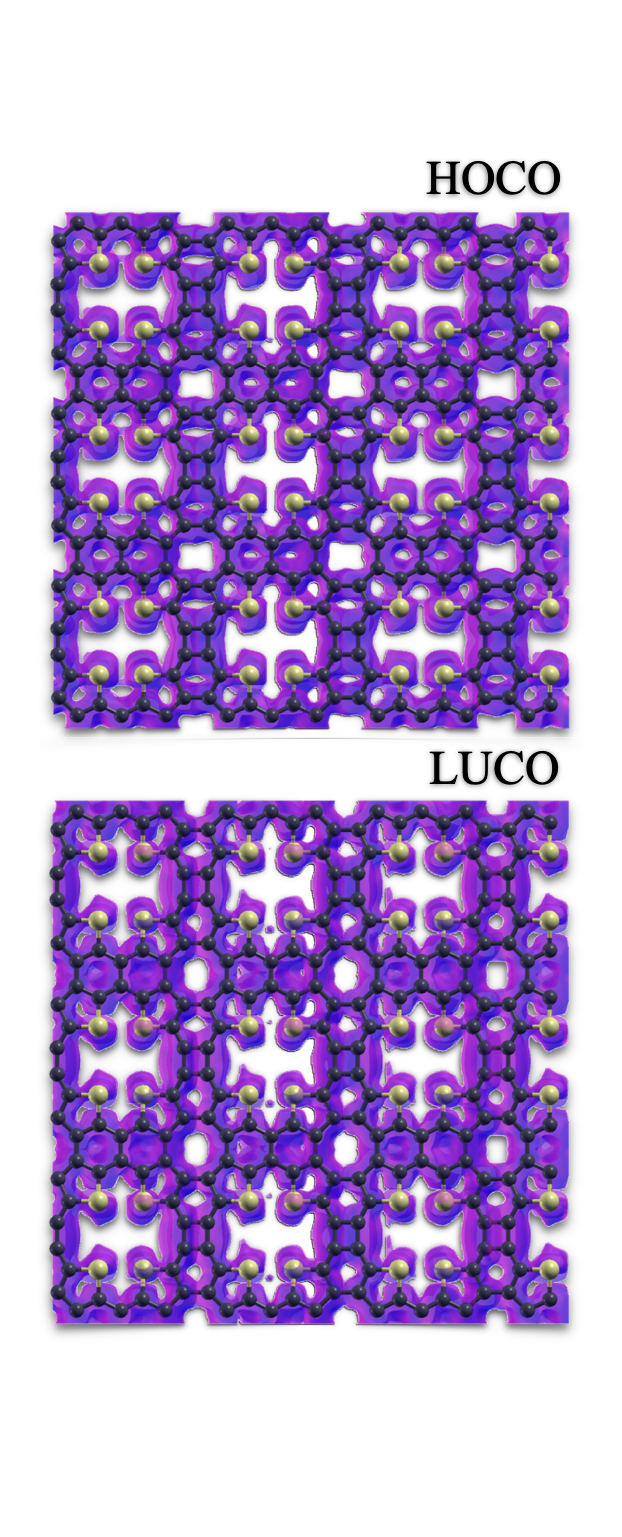}
\caption{HOCO and LUCO molecular orbitals for TTA-2D. A supercell with $104$ atoms was considered. \label{fig:orbital}}
\end{center}
\end{figure}

We Figure \ref{fig:orbital} we present the TTA-2D highest occupied crystal orbital (HOCO) and lowest unoccupied crystal orbital (LUCO). In order to 
perform this plot, we considered a post-analysis calculation with XCRYSDEN software \cite{KOKAL}. We can see that both HOCO and LUCO are well delocalized through all the structures, consistent with small bandgap systems.

\begin{figure}[]
\begin{center}
\includegraphics[width=0.83\linewidth]{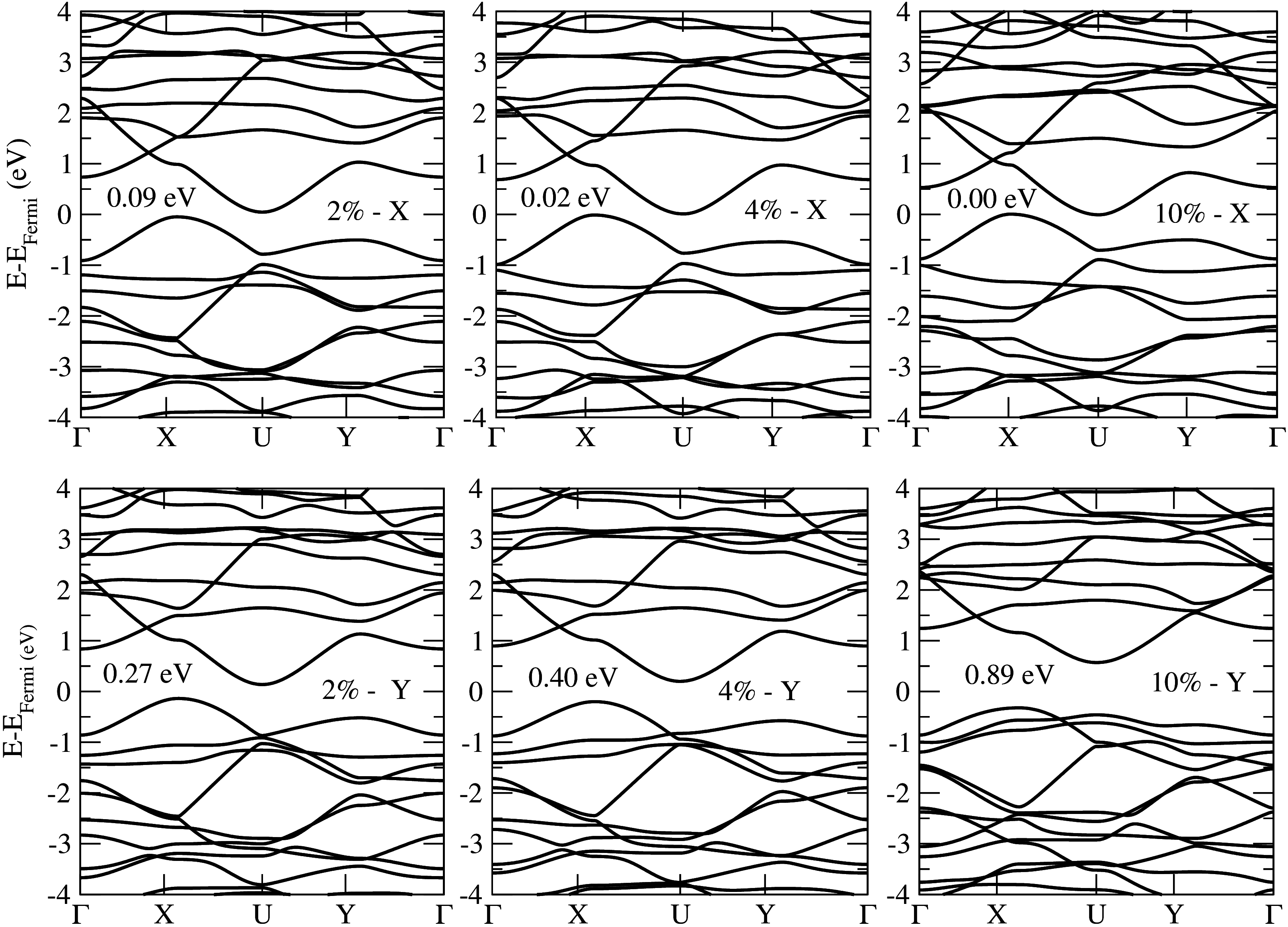}
\caption{Electronic band structure for TTA-2D under different strain levels. \label{fig:bandas_strain}}
\end{center}
\end{figure}

We also analyzed the effects of an externally applied uniaxial strain  (X and Y directions) on the electronic band structures. In Figure \ref{fig:bandas_strain} we present the results for strain values of 2\%, 4\%, and 10\% in the $X$ and $Y$ directions. We can see from the Figure that the strain effects are direction-dependent. The bandgap values decrease and increase when the strain is applied along the $X$ and $Y$ directions, respectively. This can be explained by the fact that stretching along the $X$/$Y$ directions, increases/decreases the structural planarity. The whole can be better visualized from the videos 1 and 2 in the Supplementary Materials. Increasing the planarity increases the $\pi$ character and consequently the electronic delocalization. This is extremely important for potential applications in tunable nano-electromechanical devices \cite{Deng_2018}.

\section{Optical properties}

As mentioned in the methods section, in order to perform optical calculations we used a scissor correction to produce a more precise description of the optical transitions, with electronic bandgap value based on the HSE06 functional results. That is, instead of considering the indirect bandgap value from figure \ref{fig:banpdos} (corresponding to $0.2$ eV), here the transitions were calculated considering a corrected indirect bandgap value of $0.6$ eV.

\begin{figure}[]
\begin{center}
\includegraphics[width=0.83\linewidth]{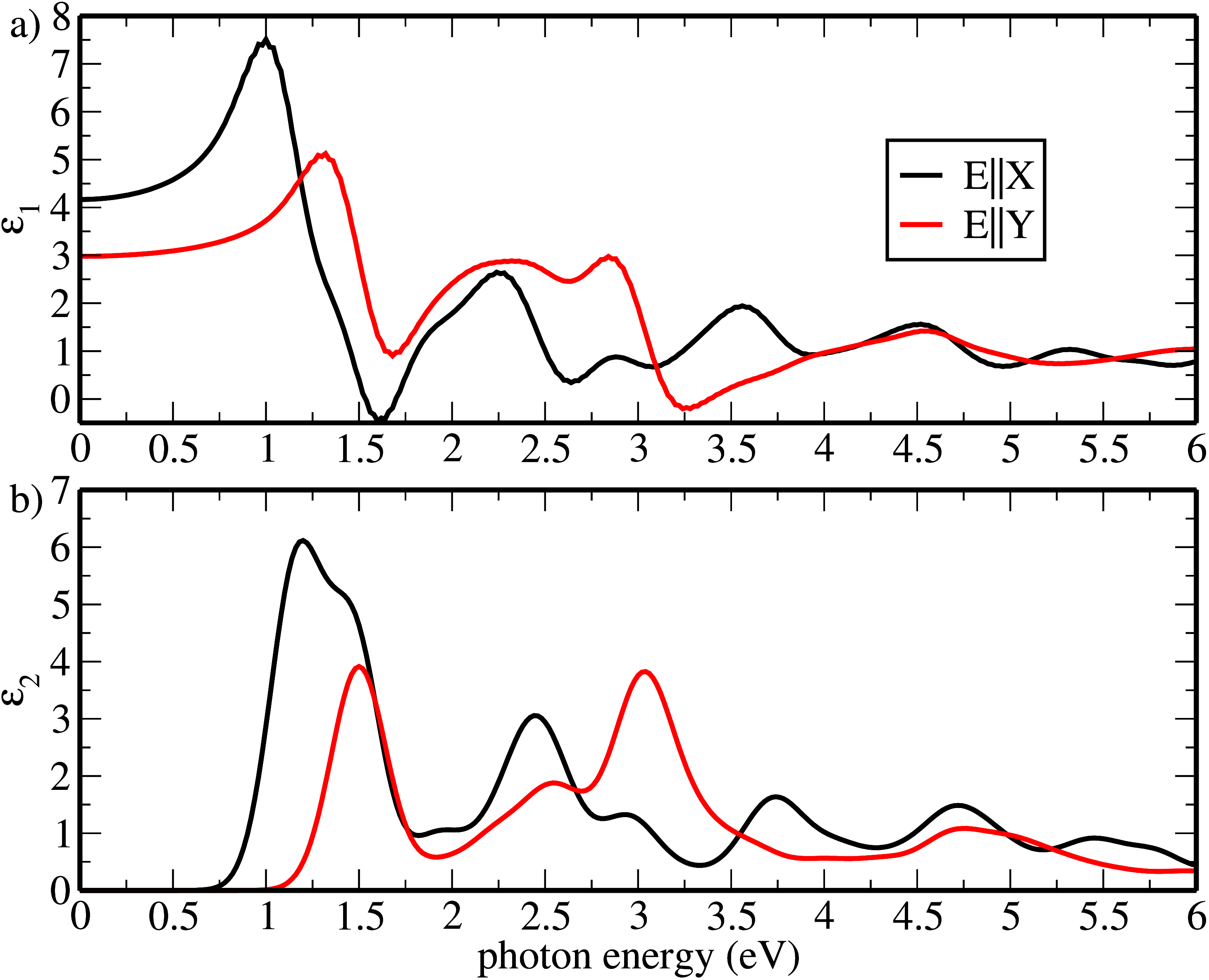}
\caption{ Real a) and imaginary b) parts of the dielectric function \textit{versus} photon energy, for an externally applied electric field polarized along the $X$ or $Y$ direction. \label{fig:epsilon}}
\end{center}
\end{figure}

In Figure \ref{fig:epsilon} we present the real a) and imaginary b) parts of the dielectric function \textit{versus} the photon energy value, considering an externally applied electric field polarized along the $X$ and $Y$ directions. We observed an anisotropic behavior for both the real and imaginary parts of the dielectric function.

The first peak in the real part, $\epsilon_1$, starts for photon energy values close to the TTA-2D bandgap value, which is $0.6$ eV, after the correction. Other peaks are related to secondary transitions due to states distant from the HOCO and LUCO ones. Optical activity is observed for photon energies up to $5.5$ eV, in the ultraviolet region. The TTA-2D static dielectric constant value can be extracted from the real part of the dielectric function in the limit where the photon energy tends to zero. Then, from figure \ref{fig:epsilon}-a, we estimated that the static dielectric constant is $4.15$ and $2.97$ for the $X$ and $Y$ directions, respectively. This information is important for the analysis of the electronic transport. The dielectric constant of the active absorbing material has an important effect on the separation of photogenerated electron-hole pairs (excitons)  \cite{exciton}.

For the imaginary part of the dielectric function, we observe in Figure \ref{fig:epsilon}-b that for the $X$ direction, the first peak starts near $0.6$ eV. This is likely related to a transition between the {\bf X} and {\bf U} symmetry points (recall the corrected bad gap is $0.6$ eV). For the $Y$ direction, we noticed that the first peak starts near $0.95$ eV. This peak is probably related to transitions from {\bf Y} to {\bf U} symmetry points. Then, we can conclude from figure \ref{fig:epsilon} that TTA-2D exhibits different optical bandgap values for the $X$ and $Y$ directions.

\begin{figure}[]
\begin{center}
\includegraphics[width=0.53\linewidth]{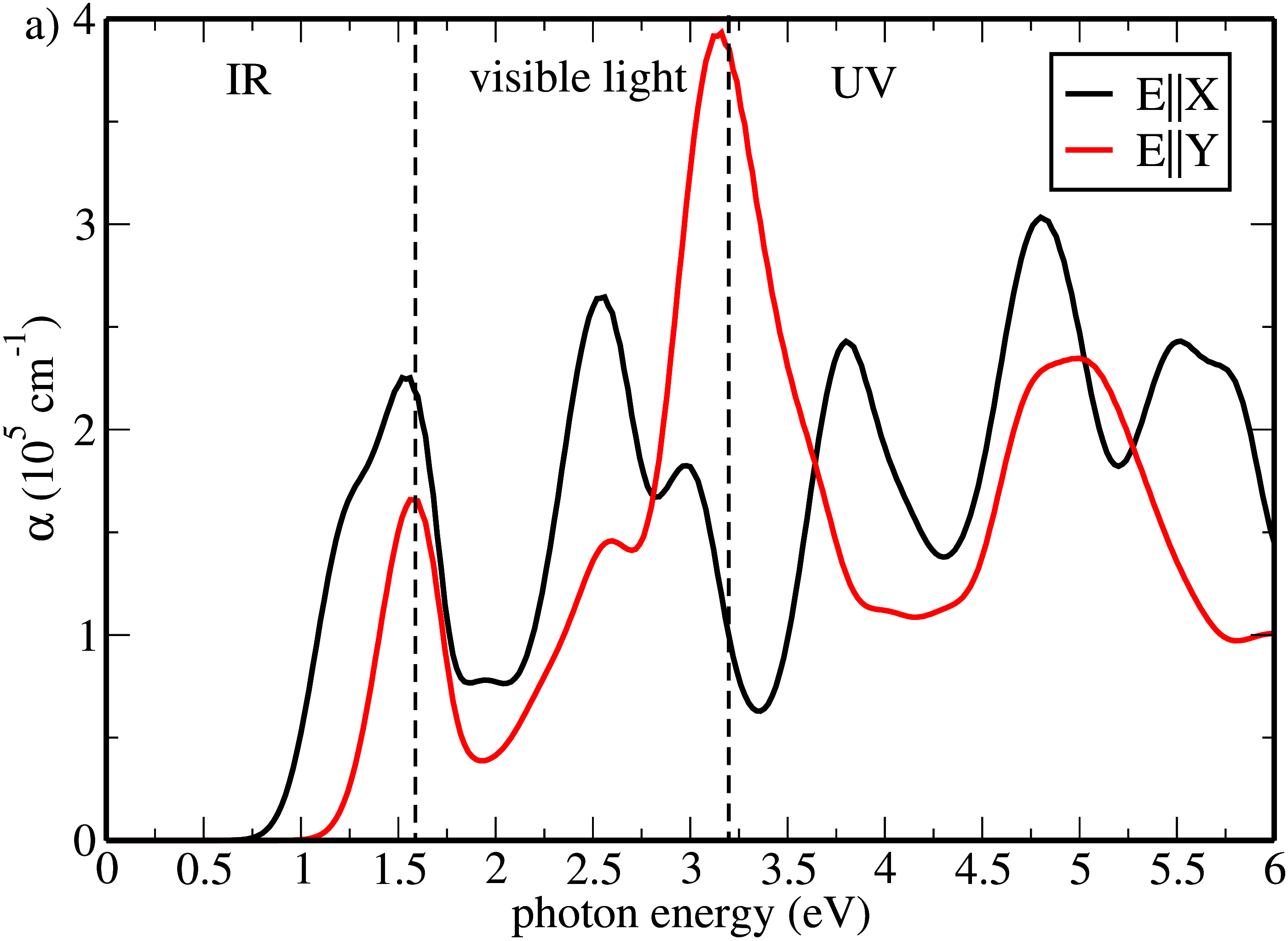}
\includegraphics[width=0.53\linewidth]{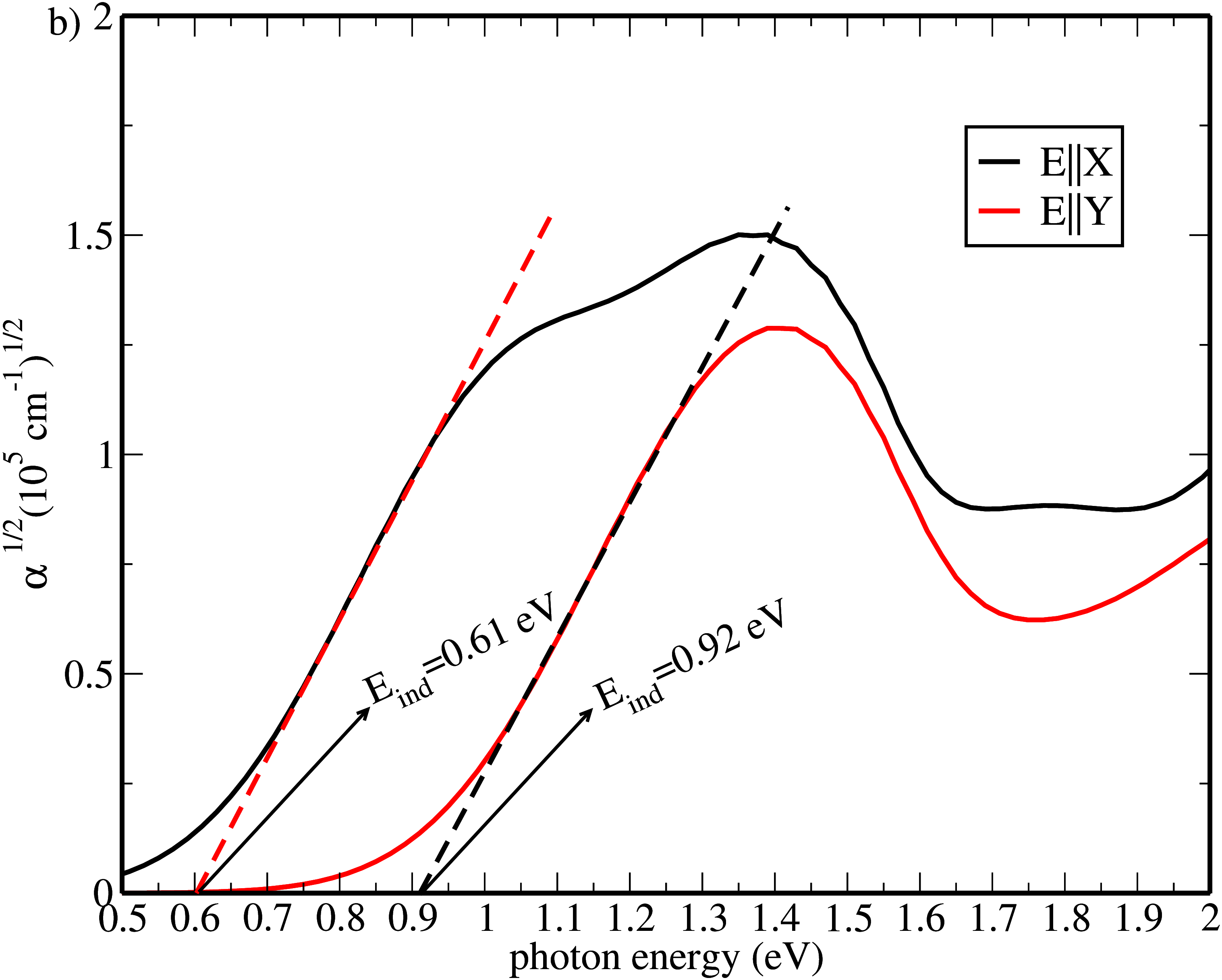}
\includegraphics[width=0.53\linewidth]{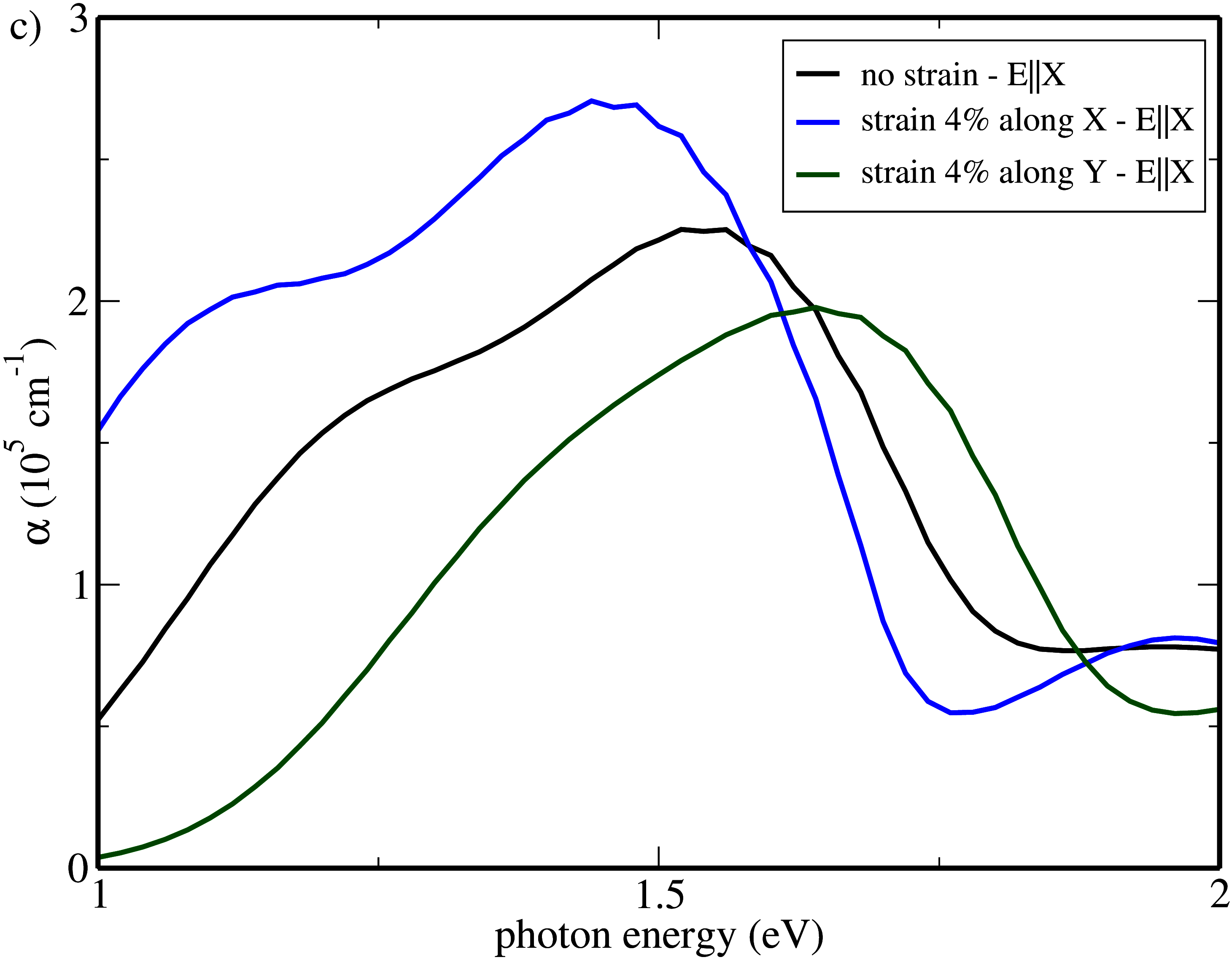}
\caption{ a) Absorption coefficient as a function of the photon energy, for an externally applied electric field polarized along the $X$ or $Y$ direction. b) Plot used to obtain the indirect optical bandgap via the {\bf Tauc's} method. c) Shift in the first absorption peak, due to the application of strain along the $X$ or $Y$ direction. \label{fig:abs}}
\end{center}
\end{figure}

In Figure \ref{fig:abs}-a we present the absorption coefficient as a function of the photon energy for an externally applied electric external field polarized along the $X$ and $Y$ directions. The absorption coefficient becomes non-zero for photon energies near $0.6$ and $0.95$ eV for $X$ and $Y$ polarization, respectively. The transitions associated with these peaks were described in the dielectric function discussions. For both polarization directions, the first peak occurs in the edge between the infrared and visible regions. In the visible, both cases present significant intensity. For a photon energy of $2.5$ eV a peak is observed for both directions, but with higher intensity for the $X$ direction. In contrast, in the edge between the visible and UV regions, the peak is more intense for the $Y$ direction. In the UV range TTA-2D has significant intensity up until a photon energy equal to 5.5 eV. Based on these results, we can conclude that TTA-2D can be used in optoelectronic devices to absorb light ranging from the infrared up to the UV region.

In order to determine the indirect optical bandgap for each direction, due to anisotropy effects, we use the well-known Tauc's method\cite{tauc}. It basically consists of a linear fit in the plot of $\alpha ^{1/2}$ as a function of the photon energy value, as displayed in Figure \ref{fig:abs}-b. The estimated indirect optical bandgap is obtained by the value where the line crosses the abscissa axis. Using this approach the obtained bandgap values were $0.61$ eV and $0.92$ eV for the $X$ and $Y$ directions, respectively. The first value is associated with an optical transition from {\bf X} to {\bf U}, and the second one with a transition from {\bf Y} to {\bf U}. These symmetry points were presented in Figure \ref{fig:banpdos}.
As discussed above, we showed that it is possible to tune the bandgad applying strain. In figure \ref{fig:abs}-c we present the results for the case of 4$\%$. We can see from this Figure that the first peak is shifted towards the infrared or the UV regions, depending on the strain direction. 

\begin{figure}[]
\begin{center}
\includegraphics[width=0.85\linewidth]{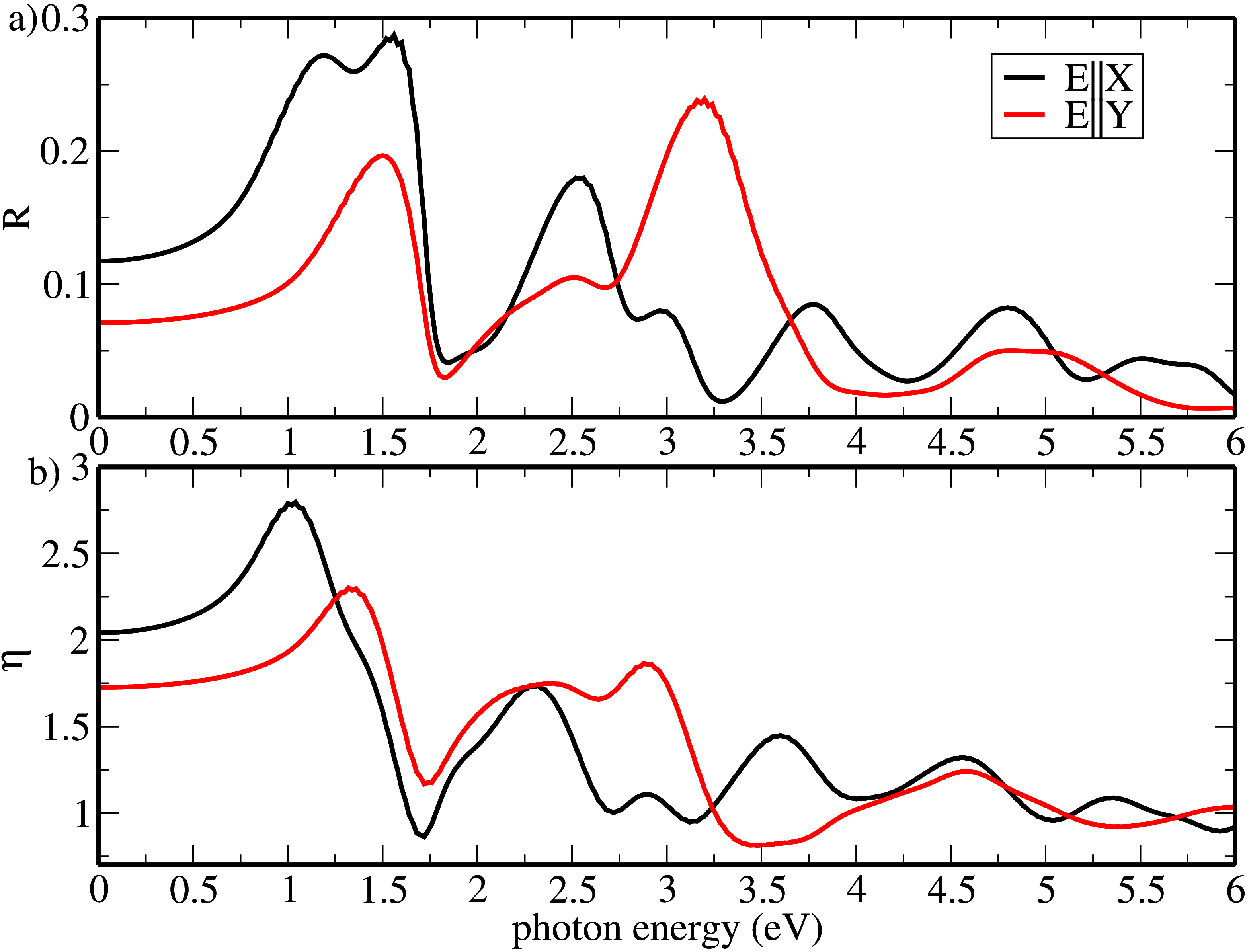}
\caption{a) Reflectivity and b) refractive  indices as a function of the photon energy. \label{fig:ref}}
\end{center}
\end{figure}

In Figure \ref{fig:ref} we present reflectivity (\ref{fig:ref}-a) and refractive (\ref{fig:ref}-b) indices, as a function of the photon energy for $X$ and $Y$ polarizations. The maximum reflectivity in the infrared occurs in a region where absorption is also significant, at energy values around $1.25-1.50$eV. However, the refractive index is also larger in this region. In the maximum 30$\%$ of infrared light is reflected, as the refractive index is larger, the tendency is that the majority of the incident light will be refracted. Then, for the light incident in TTA-2D, the majority penetrates, and only a small fraction is reflected.

This behavior is also observed in the visible and UV regions. In the visible range, the refractive index has at least one order of magnitude larger than reflectivity,  this is an important criterion for that the material can be used in solar cell applications and optical devices \cite{C3TC32252A,C5CP02938D}. In the UV region, the reflection is very small, smaller than 10$\%$, therefore the incident light tends to be refracted. In the limit of small photon energy, we can determine the static refractive index of TTA-2D, in a manner similar to what we did to obtain the static dielectric constant in Figure \ref{fig:epsilon}-a. From plot \ref{fig:ref}-b we obtained the following values for the static refraction index: $2.10$ and $1.74$, for the $X$ and $Y$ directions, respectively. Our analysis reveals that TTA-2D can be used in optoelectronics device applications with photon energies ranging from $1.5$ to $5.5$ eV, spanning from the infrared to the ultraviolet regions.

\section{Conclusions}
In summary, we used first-principles methods to investigate the structural stability as well as the mechanical, electronic, and optical properties of a new 2D structure, named Thiophene-Tetrathia-Annulene monolayer (TTA-2D).

Our results show that TTA-2D is a small indirect bandgap semiconductor ($0.2$ eV with GGA-PBE and $0.6$ eV with HSE06). This bandgap value can be reduced by applying uniaxial strain along the $X$ direction, and TTA-2D becomes metallic for an applied strain of $10\%$. Conversely, the bandgap can be increased by applying uniaxial strain along the $Y$ direction, reaching $1$ eV for an applied strain of $10\%$. Note that bandgap values for strained structures were obtained using  GGA-PBE. TTA-2D has an enthalpy of formation of $-8.6$ eV/atom, only $0.2$ eV/atom larger than graphene. The system is also structurally stable, as we did not observe negative frequencies in the phonon dispersion calculation. \textit{Ab initio} molecular dynamics simulation results show the TTA-2D is thermally stable up to $T=1000$~K. Regarding mechanical properties, the estimated Young's modulus values were $327.1$ GPa and $476.1$ GPa along the $X$ and $Y$ directions, respectively. These values are larger than those of many other 2D systems. In regard to optical properties, TTA-2D absorbs in a large range, from infrared to ultraviolet regions. Values of refractive index and reflectivity show that TTA-2D reflects only $10\%$ of the incident light in the visible region. These results show that TTA-2D is a promising material for solar cells applications. Considering that the TTA-structural unit has been already synthesized and that TTA-2D is thermally stable even at high temperatures, its synthesis is feasible with present technologies. We hope that the present work will stimulate further studies along these lines.

\section*{Acknowledgements}

This work was financed in part by the Coordenacão de Aperfeiçoamento de Pessoal de Nível Superior - Brasil (CAPES) - Finance Code 001, CNPq, and FAPESP. The authors thank the Center for Computational Engineering \& Sciences (CCES) at Unicamp for financial support through the FAPESP/CEPID Grant 2013/08293-7. LDM would also like
to thank the support of the High Performance Computing Center at UFRN (NPAD/UFRN).

\pagebreak
\section{Supplementary Material}
\section{Stress x strain curve}

\begin{figure}[]
\begin{center}
\includegraphics[width=0.83\linewidth]{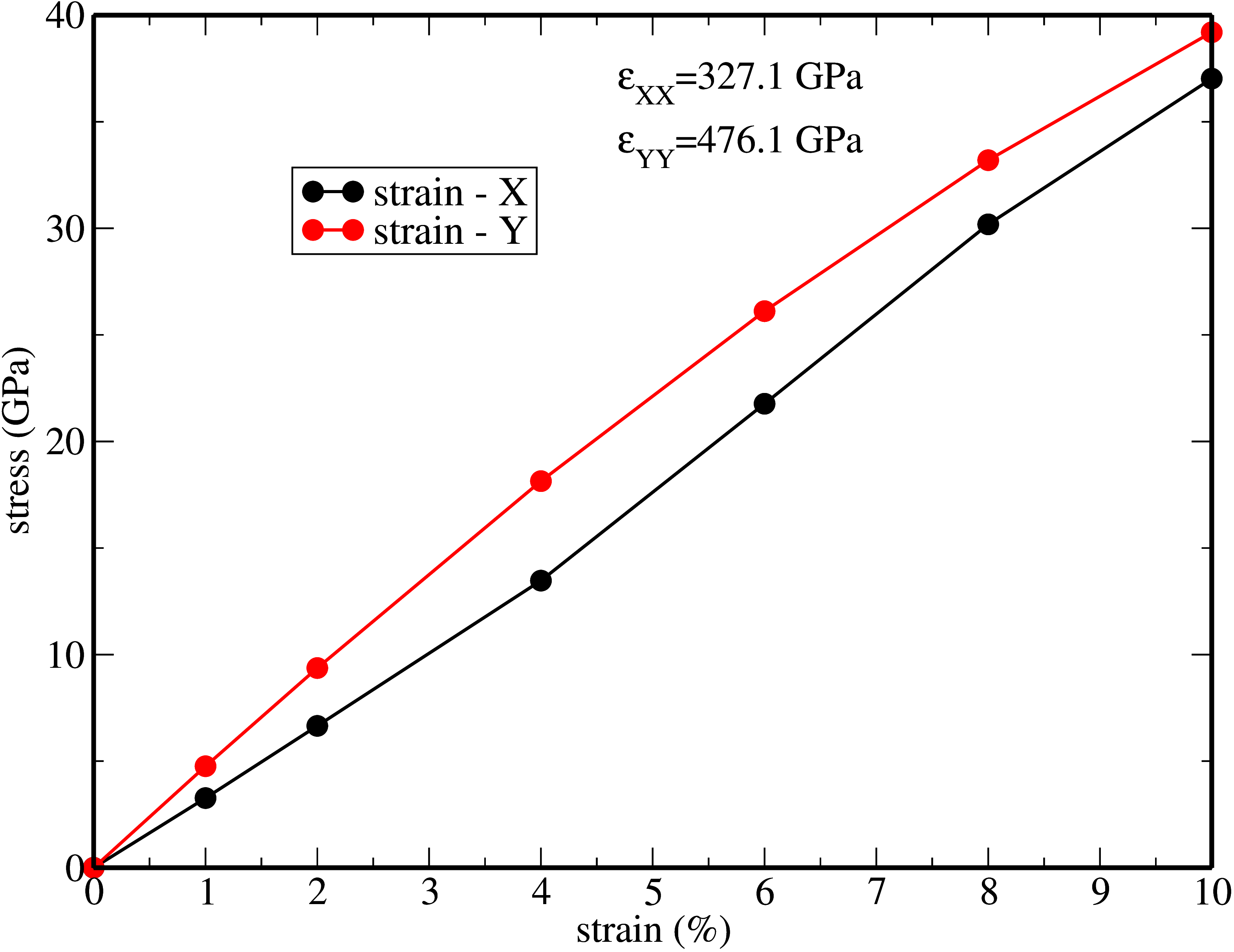}
\caption{ TTA-2D stress x strain curve for the $X$ and $Y$ directions.} \label{fig:stress}
\end{center}
\end{figure}

In Figure \ref{fig:stress} we stress-strain curves for the $X$ and $Y$ directions. The Young's modulus values were estimated from the linear coefficient through a the linear fit performed in the region with strain between $0$ and $1\%$.

\pagebreak
\bibliography{achemso-demo}

\end{document}